\documentclass[a4paper,11pt]{article}
\pdfoutput=1 

\usepackage{jheppub_modified} 

\usepackage[T1]{fontenc} 

\usepackage{verbatim}
\usepackage{color}
\usepackage{graphicx}
\usepackage{subcaption}

\usepackage{bigcenter}

\usepackage{sectsty}
\allsectionsfont{\boldmath}

\usepackage{xspace}
\usepackage[dvipsnames,table]{xcolor}
\usepackage{tabularx}
\newcolumntype{C}[1]{>{\centering\arraybackslash}p{#1}}

\newcommand{\eq}[1]{Eq.~(\ref{#1})}
\newcommand{\bib}[1]{Ref.~\cite{#1}}
\newcommand{\refs}[1]{Refs.~\cite{#1}}
\newcommand{\bibs}[1]{\cite{#1}}
\newcommand{\fig}[1]{Fig.~\ref{#1}}
\newcommand{\tab}[1]{Table~\ref{#1}}

\newcommand{\sect}[1]{Section~\ref{#1}}
\newcommand{\ssect}[1]{Subsection~\ref{#1}}

\newcommand{\bea}{\begin{eqnarray}}
\newcommand{\eea}{\end{eqnarray}}

\newcommand{\nn}{\nonumber}
\newcommand{\crn}{\nonumber \\}

\newcommand{\fr}{\frac}

\newcommand{\gev}{{\unskip\,\text{GeV}}}
\newcommand{\tev}{{\unskip\,\text{TeV}}}

\title{Doubly-polarized $WZ$ hadronic cross sections at NLO QCD+EW accuracy}

\author[a]{Duc Ninh Le,}
\author[b]{Julien Baglio}

\affiliation[a]{Faculty of Fundamental Sciences, PHENIKAA University, Hanoi 12116, Vietnam}
\affiliation[b]{Theoretical Physics Department, CERN, CH-1211 Geneva 23, Switzerland}

\emailAdd{ninh.leduc@phenikaa-uni.edu.vn}
\emailAdd{julien.baglio@cern.ch}

\preprint{CERN-TH-2022-027}      

\abstract{We present new results for next-to-leading order (NLO) electroweak (EW) corrections to 
double polarization signals in the $WZ$ production channel at the LHC using the $e^+\nu_e \mu^+ \mu^-$ 
final state. It is found that the EW corrections are most sizable in the transverse momentum distributions of the 
doubly longitudinal polarization, being around $-10$\% compared to the NLO QCD prediction at $p_{T,e}\approx 200$~GeV, which is in the accessible energy range of the current LHC data.}

\begin{document}
\maketitle
\flushbottom

\section{Introduction}
\label{sect:intro}

The CERN Large Hadron Collider (LHC) has been operating since 2009 and
has accumulated lots of data, in particular in the production of $W$ and
$Z$ electroweak (EW) gauge bosons. The detailed study of their
properties allows theorists and experimentalists for probing deeply the
Standard Model (SM) and in particular the EW symmetries, as
well as for searching for potential new-physics effects signaled by
deviations from SM expected shapes in various observables. With 13 TeV
data as well as with new data coming from run 3 and beyond in the next
years, it is possible to study non-trivial observables such as the
polarization of the gauge bosons, in particular in the four-lepton
channel via $ZZ$ production and in the three-lepton channel via $WZ$
production. The latest measurements from ATLAS and CMS collaborations
in the three-lepton channel can be found in
\refs{ATLAS:2021wob,CMS:2021icx}, respectively.

Higher order QCD and EW corrections to three-lepton production in the
$WZ$ channel have reached a high precision. The next-to-leading order
(NLO) QCD corrections were calculated in
\refs{Ohnemus:1991gb,Frixione:1992pj} for on-shell production and in
\refs{Dixon:1998py,Dixon:1999di} for off-shell production. The NLO EW corrections were presented in
\refs{Accomando:2004de,Bierweiler:2013dja,Baglio:2013toa,
  Biedermann:2017oae}, showing in particular the importance of the
quark-photon induced correction. The full NLO QCD predictions
including full off-shell and spin-correlation effects for leptonic
final states can be numerically calculated with the help of public
computer programs such as {\tt
  MCFM}~\cite{Campbell:1999ah,Campbell:2011bn} or {\tt
  VBFNLO}~\cite{Arnold:2008rz,Baglio:2014uba}. In 2018 these
calculations have been extended to include anomalous couplings effects
at the NLO QCD+EW accuracy as well~\cite{Chiesa:2018lcs}. QCD
precision has reached the next-to-next-to-leading order (NNLO)
accuracy~\cite{Gehrmann:2015ora,Grazzini:2016swo,Grazzini:2017ckn} and
a combination of NLO EW and NNLO QCD corrections has been performed in
\bib{Grazzini:2019jkl}. Parton shower effects have also been
calculated at NLO QCD~\cite{Melia:2011tj,Nason:2013ydw}, later
extended to include SM effective field theory effects in
\refs{Baglio:2019uty,Baglio:2020oqu}, while the consistent matching of
NLO QCD+EW corrections has been performed in \bib{Chiesa:2020ttl}.

As more data is available, there is a growing interest in the study of
the polarization of the gauge bosons in the three-lepton
channel \footnote{The two-lepton plus missing energy production is 
also interesting but more difficult to measure. Very recently, an NNLO QCD 
polarization study of the $W^+W^-$ production has been performed in \bib{Poncelet:2021jmj}.}. 
Notably, ATLAS presented in 2019 results for 
angular observables with 13 TeV data in the $WZ$
channel~\cite{ATLAS:2019bsc}. On the theory side, the
study of gauge boson polarizations effects started in the
eighties~\cite{Bilchak:1984gv,Willenbrock:1987xz} and the NLO QCD
corrections were included in \bib{Stirling:2012zt}. 
The EW corrections
have been calculated in detail in
\refs{Baglio:2018rcu,Baglio:2019nmc}. The latter studies have
introduced in particular the concept of fiducial polarization
observables constructed out of the final-state angular observables in
the fiducial volume, including the experimental cuts, but they have
not investigated the separation of polarization states at the
amplitude level. In order to do it is necessary to study three-lepton
production in the double-pole approximation (DPA) where the production
and decay amplitudes are calculated in the on-shell approximations,
and then combined using a sum over all polarizations retaining the
full phase-space in the gauge boson propagators. This study has been
performed at NLO QCD in \bib{Denner:2020eck}, but is still lacking the
NLO EW corrections, contrary to the four-lepton
channel~\cite{Denner:2021csi}. Our study closes the gap by including
the NLO QCD and EW corrections in the DPA for the three-lepton
channels, separating the polarization states at the amplitude
level. In this letter, we provide results for the $W^+ Z$ channel using the 
same fiducial cuts and reference frame as ATLAS~\cite{ATLAS:2019bsc}.

The paper is organized as follows. The definition of
polarizations and a sketch of our calculation framework 
are given in \sect{sect:cal}. Numerical results at the 13 TeV LHC are presented in 
\sect{sect:res}, starting with the integrated polarized cross sections in
\ssect{sect:XS} before describing kinematical
distributions in \ssect{sect:dist}. Conclusions are provided in \sect{sect:conclusion}. 

\section{Calculation of polarized cross sections}
\label{sect:cal}
The process considered in this paper reads
\bea
p + p \to V_1(q_1) + V_2(q_2) \to \ell_1(k_1) + \ell_2(k_2) + \ell_3
(k_3) + \ell_4(k_4) + X,
\label{eq:proc1_VV}
\eea
where the final-state leptons can be either $e^+\nu_e\mu^+\mu^-$ or 
$e^-\bar{\nu}_e\mu^+\mu^-$ and the intermediate gauge bosons are $V_1=W^\pm$, $V_2=Z$. 

\begin{figure}[h!]
  \centering
  \includegraphics[width=0.4\textwidth]{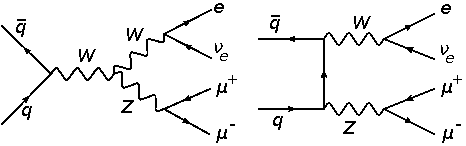}
  \caption{Doubly resonant diagrams at leading order.}
  \label{fig:LO_diags_double}
\end{figure}

\begin{figure}[h!]
  \centering
  \includegraphics[width=0.4\textwidth]{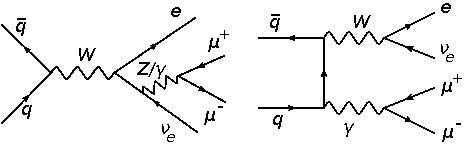}
  \caption{Non-doubly resonant diagrams at leading order.}
  \label{fig:LO_diags_NonDouble}
\end{figure}

The polarization signals are defined using the double-pole approximation. 
In this framework, the final state leptons are created from intermediate states of an on-shell 
diboson system as can be seen from \fig{fig:LO_diags_double}. 
Non-double-pole contributions such as $W \to 4l$ or $W\gamma \to 4l$ shown in \fig{fig:LO_diags_NonDouble} are excluded.  


Each massive gauge boson has three polarization states, two transverse (T) and one longitudinal (L). The diboson system has 
therefore nine polarization states. One can thus imagine the process \eq{eq:proc1_VV} occurs in a way similar to 
a 9-slit experiment, each slit corresponds to a polarization state of the $WZ$ system. It is therefore natural to expect that 
there must be interferences between waves passing through the different slits. 

Quantitatively, the contributions of those nine polarization states can be calculated as follows. 
At LO, the amplitude in the DPA is defined as (see
e.g. \bib{Denner:2000bj})
\bea
\mathcal{A}_\text{LO,DPA}^{\bar{q}q'\to V_1V_2\to 4l} = \fr{1}{Q_1Q_2}
\sum_{\lambda_1,\lambda_2=1}^{3}
\mathcal{A}_\text{LO}^{\bar{q}q'\to V_1V_2}\mathcal{A}_\text{LO}^{V_1\to
    \ell_1\ell_2}\mathcal{A}_\text{LO}^{V_2\to \ell_3\ell_4}
,\label{eq:LO_DPA}
\eea
with 
\bea
Q_j = q_j^2 - M_{V_j}^2 + iM_{V_j}\Gamma_{V_j},
\label{eq:Qi_def}
\eea
where $q_1 = k_1+k_2$, $q_2 = k_3 + k_4$, $M_V$ and $\Gamma_V$ are the physical mass and width 
of the gauge bosons. We note that all helicity amplitudes in the numerator must be calculated
using on-shell momenta. This is important to make sure that the amplitudes 
are gauge invariant. The OS momenta can be calculated from the original momenta $k_i$ 
by means of an OS mapping. This mapping is not unique. However, it has been pointed out in
\bib{Denner:2000bj} that different mappings lead to small differences
of the order of $\alpha\Gamma_V/(\pi M_V)$.

From \eq{eq:LO_DPA} we can define the nine polarization contributions and their interferences. 
For example, the longitudinal-longitudinal (LL) contribution is calculated by selecting the 
$\lambda_1=\lambda_2=2$ term in the r.h.s. Similarly, the transverse-transverse (TT) polarization is 
obtained by adding the $(1,1)$, $(1,3)$, $(3,1)$, $(3,3)$ terms. Interferences between these polarization states 
are therefore included in the TT contribution. In the following, we will classify all polarization states into four groups 
LL, LT, TL, TT. The unpolarized result, calculated from \eq{eq:LO_DPA}, is thus the sum of these contributions and their interferences. 
While the unpolarized cross section is Lorentz invariant, 
individual polarized cross sections are not, hence dependent on a chosen reference frame. 
In this paper, we provide results in the $WZ$ center-of-mass system (c.m.s.), which was recently used by ATLAS in 
\bib{ATLAS:2019bsc}.  

NLO QCD and EW corrections are also calculated in the DPA. The NLO QCD calculation has been done in \cite{Denner:2020eck}, 
which is the same as for the $WW$ \cite{Denner:2020bcz} and $ZZ$ \cite{Denner:2021csi} production. 
NLO EW corrections for the $ZZ$ case has been very recently calculated in \cite{Denner:2021csi}. For the present process of $WZ$, the NLO EW corrections are more complicated because the photon can be radiated off the $W$ boson, which is treated as on-shell. Technical details of this calculation will be provided in a separate longer publication \cite{WZ_long}. Concerning the OS mappings, the mappings $\text{DPA}^{(2,2)}$ and $\text{DPA}^{(3,2)}$ given in \cite{Denner:2021csi} for $1\to 2$ and $1\to 3$ decays of the massive gauge bosons, repectively, are used for both gauge bosons.  

\section{Numerical results}
\label{sect:res}
The input parameters are the same as in \bib{Baglio:2018rcu}. We re-provide them here for the 
sake of completeness. 
\begin{eqnarray}
G_{\mu} = 1.16637\times 10^{-5} \gev^{-2}, \,
M_W=80.385 \gev, \, 
M_Z = 91.1876 \gev, \crn 
\Gamma_W = 2.085\gev, \, \Gamma_Z = 2.4952\gev, \, 
M_t = 173 \gev, \, M_H=125\gev.
\label{eq:param-setup}
\end{eqnarray}
The masses of the leptons and the light quarks, {\it i.e.} all but the top
mass, are neglected. The electromagnetic coupling is
calculated as $\alpha_{G_\mu}=\sqrt{2}G_\mu
M_W^2(1-M_W^2/M_Z^2)/\pi$. For the factorization and renormalization scales, we use 
$\mu_F = \mu_R = (M_W + M_Z)/2$. Moreover, the parton distribution functions (PDF) are 
calculated using the Hessian set \\
{\tt
  LUXqed17\char`_plus\char`_PDF4LHC15\char`_nnlo\char`_30}~\bibs{Manohar:2016nzj,Manohar:2017eqh,Butterworth:2015oua,Dulat:2015mca,Harland-Lang:2014zoa,Ball:2014uwa,Gao:2013bia,Carrazza:2015aoa,Watt:2012tq,deFlorian:2015ujt} via the library {\tt LHAPDF6}~\bibs{Buckley:2014ana}.

We will present results for the LHC at a center-of-mass energy
$\sqrt{s} = 13\,\tev$. The extra
parton radiation occurring in the NLO QCD corrections is treated 
inclusively and no jet cuts are applied. Lepton-photon
recombination is implemented, where the momentum of a given charged
lepton $\ell$ is redefined as $p'_\ell = p_\ell + p_\gamma$ if $\Delta
R(\ell,\gamma) \equiv \sqrt{(\Delta\eta)^2+(\Delta\phi)^2}< 0.1$. The letter
$\ell$ denotes either $e$ or $\mu$. After the possible lepton-photon recombination 
we then apply the following phase-space cut:
\bea
        p_{T,\mu} > 15\gev, \quad p_{T,e} > 20\gev, \quad 
        |\eta_\ell|<2.5,\crn
        \Delta R\left(e,\mu^\pm\right) > 0.3, \quad \Delta
        R\left(\mu^+,\mu^-\right) > 0.2,\label{eq:cut_default}\\
        \left|m_{\mu^+\mu^-} - M_Z\right| < 10\gev, \quad m_{T,W} >
        30\gev,\nn
\eea
which is used by ATLAS in \refs{Aaboud:2016yus,ATLAS:2019bsc} to define the
fiducial phase space. 

\subsection{Integrated polarized cross sections}
\label{sect:XS}
We first present results for the doubly polarized integrated cross
sections in \tab{tab:xs_fr}. The unpolarized value, which is the sum
of the polarized ones and their interference (last row), is also
provided. For the unpolarized cross section, the NLO QCD corrections
are rather large, of the order of +80\%, while the NLO EW corrections
(usually denoted by $\delta_{\text{EW}}$ in the literature) are
negative and amount to -4.2\%. We define a correction factor
$\bar{\delta}_{\text{EW}}$ which gives the amount of NLO EW
corrections with respect to the NLO QCD cross section, so that we can
assess the importance of the EW corrections with respect to the
QCD-corrected cross sections. For the
unpolarized cross section we get
$\bar{\delta}_{\text{EW}}=-2.3\%$. We also provide the three-point scale
uncertainty, obtained by comparing the results obtained with $\mu_F^{} =
\mu_R^{} = \mu_0^{} = (M_W^{}+M_Z^{})/2$ with those obtained using
$\mu_F^{}=\mu_R^{} = 2\mu_0^{}$ and with $\mu_F^{}=\mu_R^{}=\mu_0^{}/2
$. The LO (and NLO EW) scale uncertainty of the
unpolarized cross section is quite small, $\sim +5\%/-6\%$, while the
NLO QCD (and NLO QCD+EW, written also as NLO QCDEW) scale uncertainty is slightly smaller, $\sim
+4.5\%/-3.5\%$.

We have computed the polarized cross sections for the four
polarization combinations: the doubly longitudinal polarization
$W_L^{}Z_L^{}$, the doubly transverse polarization $W_T^{}Z_T^{}$, as
well as the mixed polarizations $W_L^{}Z_T^{}$ and $W_T^{}Z_L^{}$. We
also provide numbers for the interference term, that when summed with
the four polarized cross sections helps to recover the unpolarized
cross section. Both at LO and at NLO the doubly transverse
polarization cross section has the highest fraction, around 70.5\% at
LO and 63\% at NLO QCDEW. The NLO EW corrections are quite small and
negative, as in the unpolarized case, and of the order of -5\% while
$\bar{\delta}_{\text{EW}}=-3\%$. There is also a slight reduction of the
scale uncertainty for the doubly transverse cross section, from
+4.5\%/-5.6\% at LO down to +4\%/-3\% at NLO QCDEW.

The doubly longitudinal polarization contributes to 8\% to the
unpolarized cross section at LO and to 5.6\% at NLO QCDEW. The NLO QCD
corrections are much smaller than those of the unpolarized and doubly
transverse polarization cross sections, of the order of +30\%, while
the NLO EW corrections and $\bar{\delta}_{\text{EW}}$ are quite
similar, -4.3\% and -3.3\% respectively. There is a strong reduction
of the scale uncertainty from LO to NLO QCDEW, with +5.1\%/-6.3\% at
LO down to +1.2\%/-0.6\% at NLO QCDEW.

\begin{table}[t!]
 \renewcommand{\arraystretch}{1.3}
\begin{bigcenter}
\setlength\tabcolsep{0.03cm}
\fontsize{7.0}{7.0}
\begin{tabular}{|c|c|c|c|c|c|c|c|c|}\hline
  & $\sigma_\text{LO}\,\text{[fb]}$ & $f_\text{LO}\,\text{[\%]}$  & $\sigma^\text{EW}_\text{NLO}\,\text{[fb]}$ & $f^\text{EW}_\text{NLO}\,\text{[\%]}$ & $\sigma^\text{QCD}_\text{NLO}\,\text{[fb]}$ & $f^\text{QCD}_\text{NLO}\,\text{[\%]}$ & $\sigma^\text{QCDEW}_\text{NLO}\,\text{[fb]}$ & $f^\text{QCDEW}_\text{NLO}\,\text{[\%]}$\\
\hline
{\fontsize{6.0}{6.0}$\text{Unpolarized}$} & $18.934(1)^{+4.8\%}_{-5.9\%}$ & $100$ & $18.138(1)^{+4.9\%}_{-6.0\%}$ & $100$ & $34.071(2)^{+4.3\%}_{-3.4\%}$ & $100$ & $33.275(2)^{+4.5\%}_{-3.6\%}$ & $100$\\
\hline
{\fontsize{6.0}{6.0}$W^+_{L}Z_{L}$} & $1.492^{+5.1\%}_{-6.3\%}$ & $7.9$ & $1.428^{+5.2\%}_{-6.4\%}$ & $7.9$ & $1.938^{+1.0\%}_{-0.5\%}$ & $5.7$ & $1.874^{+1.2\%}_{-0.6\%}$ & $5.6$\\
{\fontsize{6.0}{6.0}$W^+_{L}Z_{T}$} & $2.018^{+5.8\%}_{-7.0\%}$ & $10.7$ & $1.951^{+5.8\%}_{-7.0\%}$ & $10.8$ & $5.273^{+6.2\%}_{-5.2\%}$ & $15.5$ & $5.207^{+6.4\%}_{-5.3\%}$ & $15.6$\\
{\fontsize{6.0}{6.0}$W^+_{T}Z_{L}$} & $1.903^{+5.7\%}_{-6.9\%}$ & $10.1$ & $1.893^{+5.7\%}_{-6.9\%}$ & $10.4$ & $5.024^{+6.3\%}_{-5.3\%}$ & $14.7$ & $5.013^{+6.3\%}_{-5.3\%}$ & $15.1$\\
{\fontsize{6.0}{6.0}$W^+_{T}Z_{T}$} & $13.376^{+4.5\%}_{-5.6\%}$ & $70.6$ & $12.728(1)^{+4.6\%}_{-5.7\%}$ & $70.2$ & $21.626(2)^{+3.7\%}_{-2.8\%}$ & $63.5$ & $20.977(2)^{+4.0\%}_{-3.0\%}$ & $63.0$\\
\hline
{\fontsize{6.0}{6.0}$\text{Interference}$} & $0.144(1)^{+3.4\%}_{-4.6\%}$ & $0.8$ & $0.138(1)^{+3.3\%}_{-5.6\%}$ & $0.8$ & $0.210(3)^{+1.1\%}_{-1.5\%}$ & $0.6$ & $0.204(3)^{+0.8\%}_{-1.6\%}$ & $0.6$\\
\hline
\end{tabular}
\caption{\small Unpolarized and doubly polarized cross sections in fb
  together with polarization fractions calculated at LO, NLO EW, NLO
  QCD, and NLO QCD+EW, all in the DPA, in the $WZ$ center-of-mass system for the
  process $p p \to W^+ Z\to e^+ \nu_e \mu^+\mu^-$.   
  The statistical uncertainties (in parenthesis) are given on the last
  digits of the central prediction when significant. Three-point scale
  uncertainty is also provided for the cross sections as sub- and
  superscripts in percent.}
\label{tab:xs_fr}
\end{bigcenter}
\end{table}

The mixed polarizations contribute to around 10\% each to the
unpolarized cross section at LO and to around 15\% each at NLO QCDEW. The
NLO QCD corrections are much bigger than in the other polarizations:
the ratio NLO QCD/LO amounts to around 2.6. The NLO EW corrections are
very small in comparison as we get $\bar{\delta}_{\text{EW}}=-1.3\%$
for the $W_L^{}Z_T^{}$ cross section and $\bar{\delta}_{\text{EW}}=-0.2\%$
for the $W_T^{}Z_L^{}$ cross section. The scale uncertainty is quite
similar at LO and NLO QCDEW.

The last row of \tab{tab:xs_fr} gives the results for the interference
term. It is one to two orders of magnitude smaller than the doubly
polarized cross sections and contributes to only 1\% to the
unpolarized cross section at LO and to 0.6\% at NLO QCDEW, indicating
that the interference effects are subdominant. The scale uncertainty
at NLO EW is slightly bigger than at LO, but given that the cross section is so
small this may be attibuted to numerical effects: we calculate the
interference cross section as the difference between the unpolarized
cross section and the sum of the doubly-polarized cross sections, so
that the scale variation is very sensitive to the numerical error on
the cross sections. This effect is mitigated when comparing NLO QCD
and NLO QCDEW results.

\subsection{Kinematic distributions}
\label{sect:dist}
\begin{figure}[t!]
  \centering
  \begin{tabular}{cc}
  \includegraphics[width=0.48\textwidth]{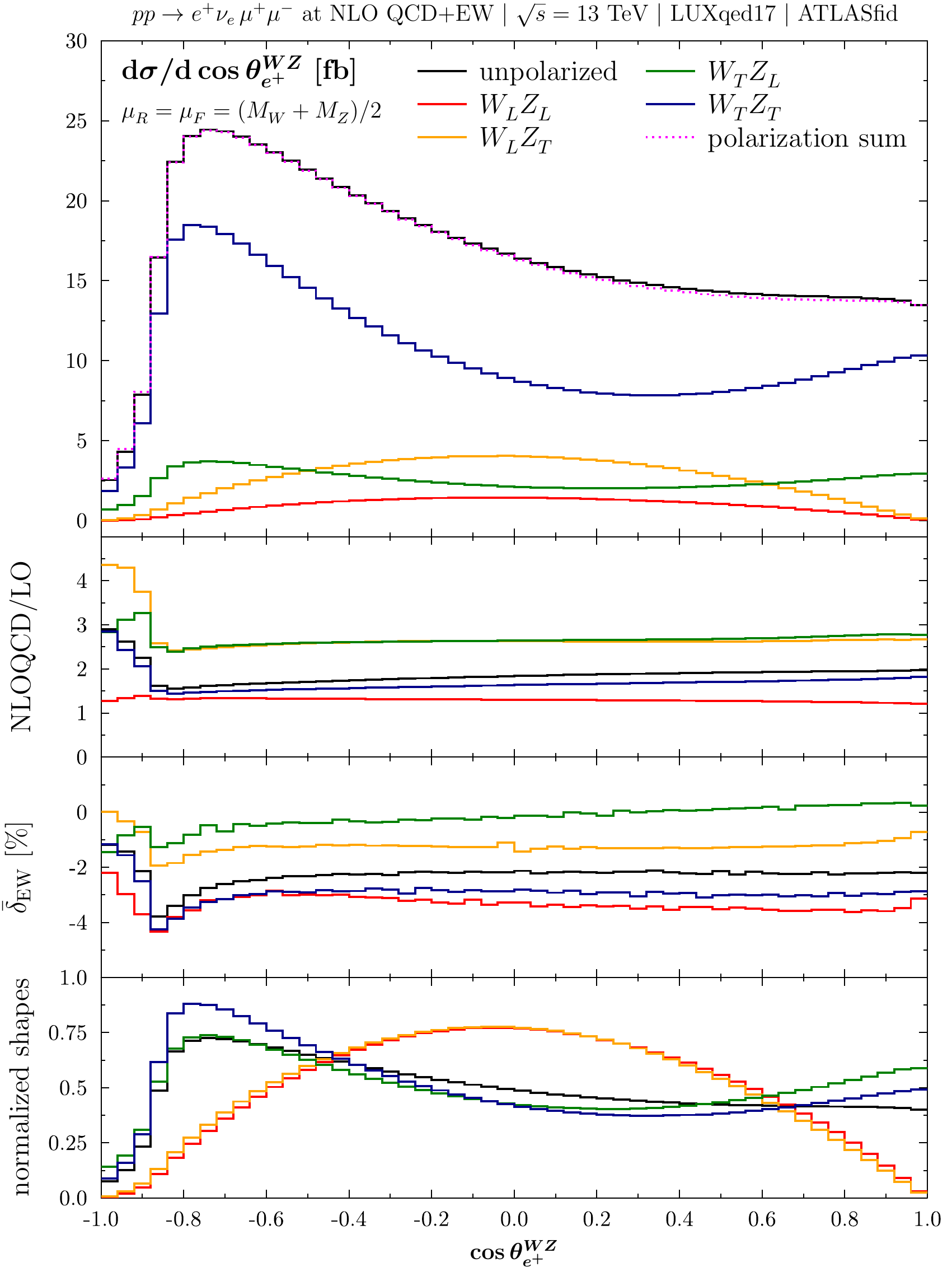}& 
  \includegraphics[width=0.48\textwidth]{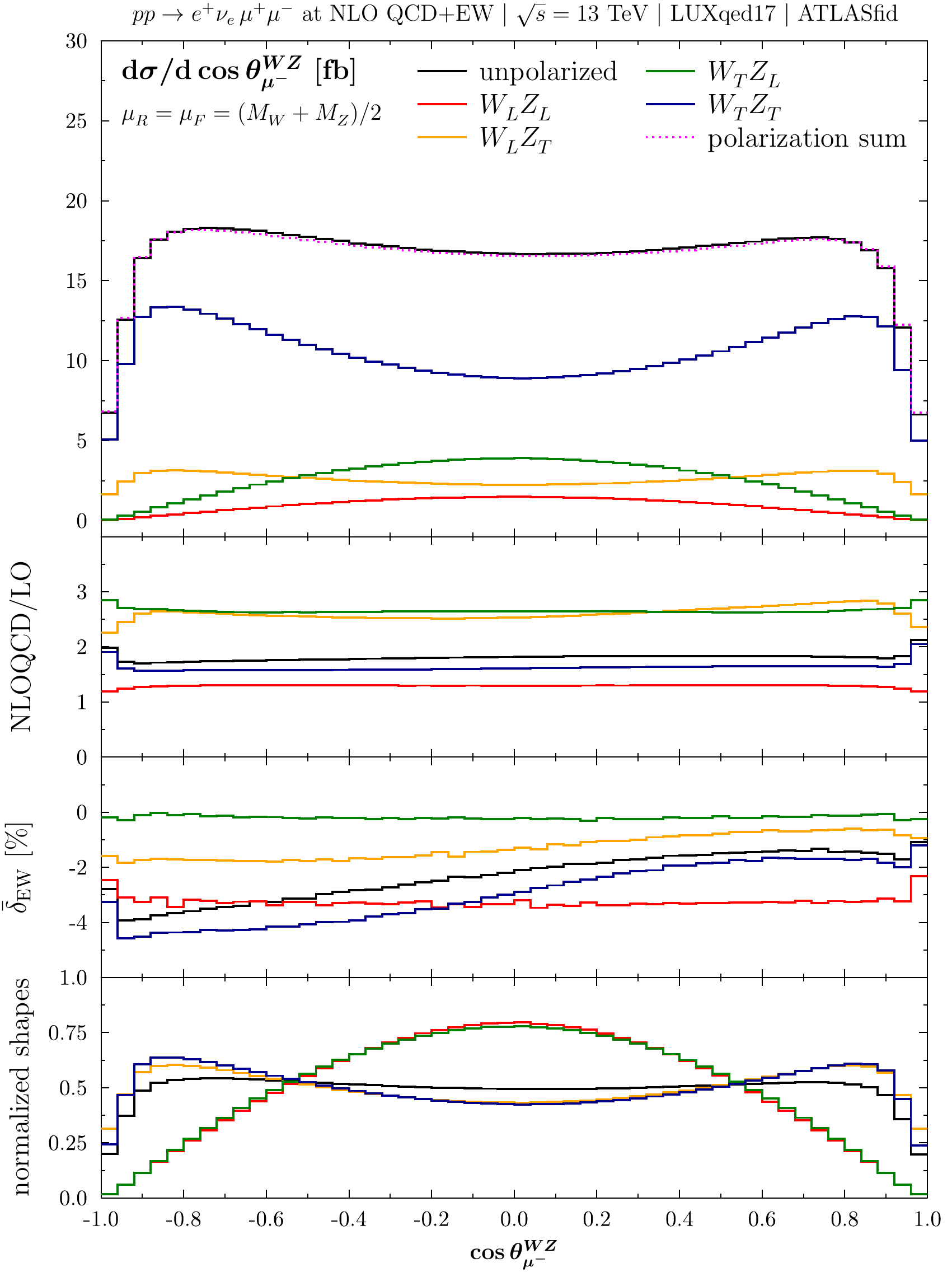}
  \end{tabular}
  \caption{Distributions in $\cos\theta^{WZ}_{e^+}$ (left) and 
    $\cos\theta^{WZ}_{\mu^-}$ (right). These angles are calculated in the $WZ$ center-of-mass system 
    (more details are provided in the text), hence 
    denoted with the $WZ$ superscript. 
    The big panel shows the absolute values of 
    the cross sections at NLO QCD+EW. 
    The middle-up panel displays the ratio of the NLO QCD cross sections to the
    corresponding LO ones. The middle-down panel shows $\bar{\delta}_{\text{EW}}$, the EW corrections relative to the NLO QCD 
    cross sections, in percent. In the bottom panel, the normalized shapes of the distributions are plotted to highlight differences in shape.}
  \label{fig:dist_costheta}
\end{figure}
We now present results for differential cross sections. The most important distribution in the analysis of $W$/$Z$ boson polarizations 
is the angular distribution of the decay lepton. In this paper, the polarizations are calculated in the $WZ$ center-of-mass system. 
The charged lepton angle $\theta^{WZ}_{\ell}$ is therefore defined as the angle between the momentum of the parent gauge boson 
calculated in the $WZ$ c.m.s. ($\vec{p}^\text{WZ-cms}_V$) and the momentum of the lepton calculated in the gauge boson rest frame ($\vec{p}^\text{V-rest}_{\ell}$). From this distribution, the polarization fractions of the gauge boson can be directly extracted (see e.g. \refs{Bern:2011ie,Baglio:2018rcu}). It is therefore important to see the various sub-contributions to this distribution from the 
individual polarizations of the $WZ$ system. This information is shown in \fig{fig:dist_costheta} for the cases of $e^+$ (coming from the decay of the $W^+$ boson) and $\mu^-$ (coming from the decay of the $Z$ boson). 
The NLO QCD results have been recently presented in \cite{Denner:2020eck}, which agree very well with our results (see the middle-up panels). 
The new results of this work are the EW corrections, shown in the middle-down panel. 
We remind that the EW corrections are defined with respect to the NLO QCD results. 
We see that the EW corrections are ranging from $-4.5$\% to $+0.5$\% for both cases and for all individual double polarizations. 
The interference effects can be seen from the difference between the unpolarized cross section and the sum of the $W_T^{}Z_T^{}$, $W_T^{}Z_L^{}$, $W_L^{}Z_T^{}$, and $W_L^{}Z_L^{}$ ones. We observe that this effect is uniformly very small here. 
In the bottom panel, we highlight the shape differences by showing the normalized distributions, i.e. the distributions in the 
big panels are normalized by the corresponding integrated cross sections. 
Looking at these normalized shapes, we see that, as expected, the electron-angle distribution is insensitive to 
the polarizations of the $Z$ boson, while it is highly sensitive to the polarizations of the $W$ boson. 
The unpolarized shape is mostly defined by the $W$'s transverse polarization. 
The same things can be said for the muon case. However, the unpolarized shape is more affected by the $Z$'s
longitudial polarization. We observe also that the shapes of $W_T^{}Z_T^{}$ (blue) and $W_L^{}Z_T^{}$ (orange) are more identical than 
the electron plot (see the $W_T^{}Z_T^{}$ and $W_T^{}Z_L^{}$). In other words, the $Z_L$ and $Z_T$ are affecting the electron angle 
in a significantly different way when $|\cos\theta^{WZ}_{e^+}|\approx 1$, while the $W_L$ and $W_T$ 
are affecting the muon angle in the same manner. 
The results in \cite{Denner:2020eck} show the same behavior. 
This is rather unexpected. In \cite{Denner:2020eck}, it is attributed to 
the differences in the kinematic cuts applied on the $Z$ and 
$W$ decay leptons.
\begin{figure}[t!]
  \centering
  \begin{tabular}{cc}
  \includegraphics[width=0.48\textwidth]{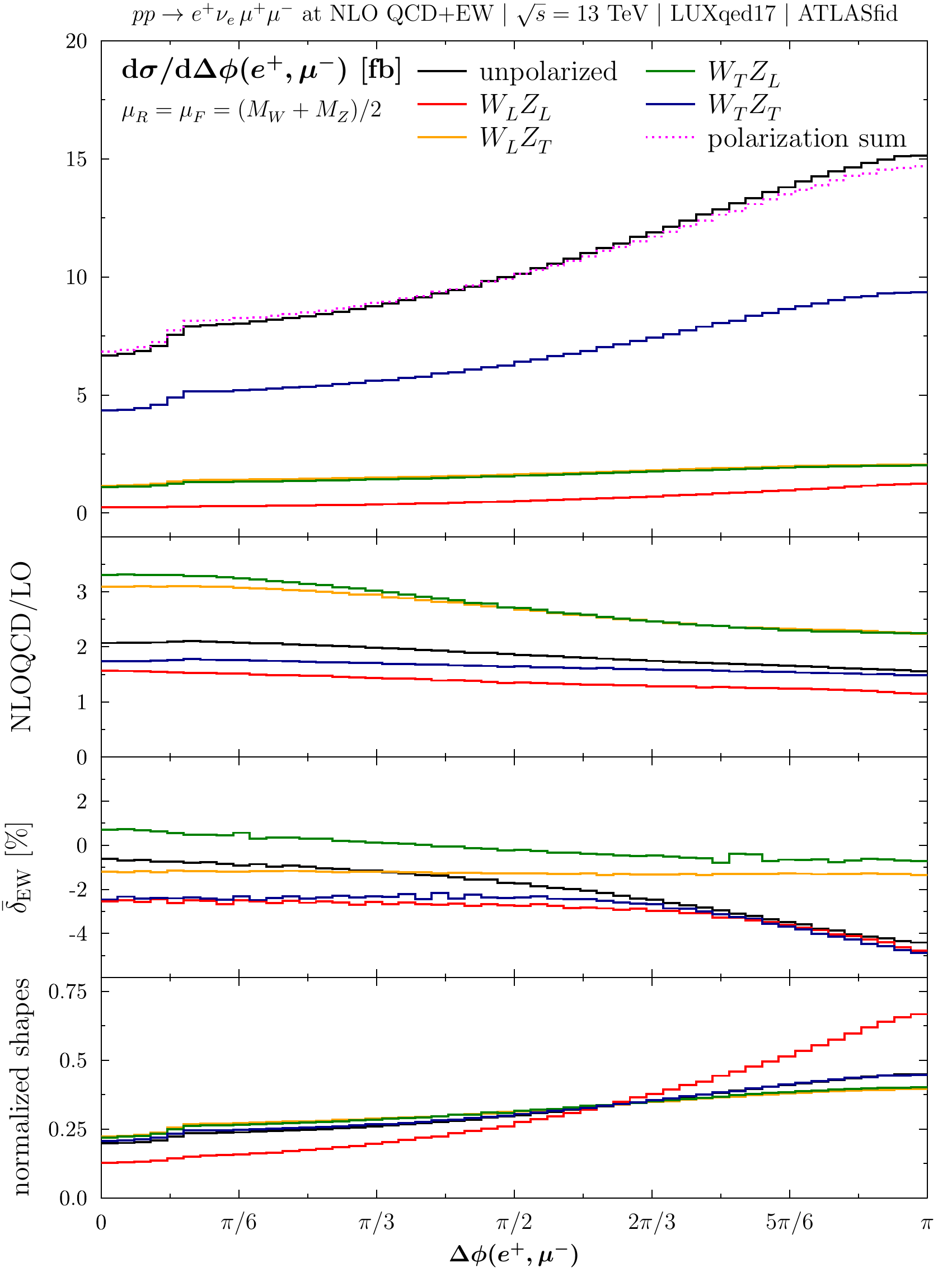}& 
  \includegraphics[width=0.48\textwidth]{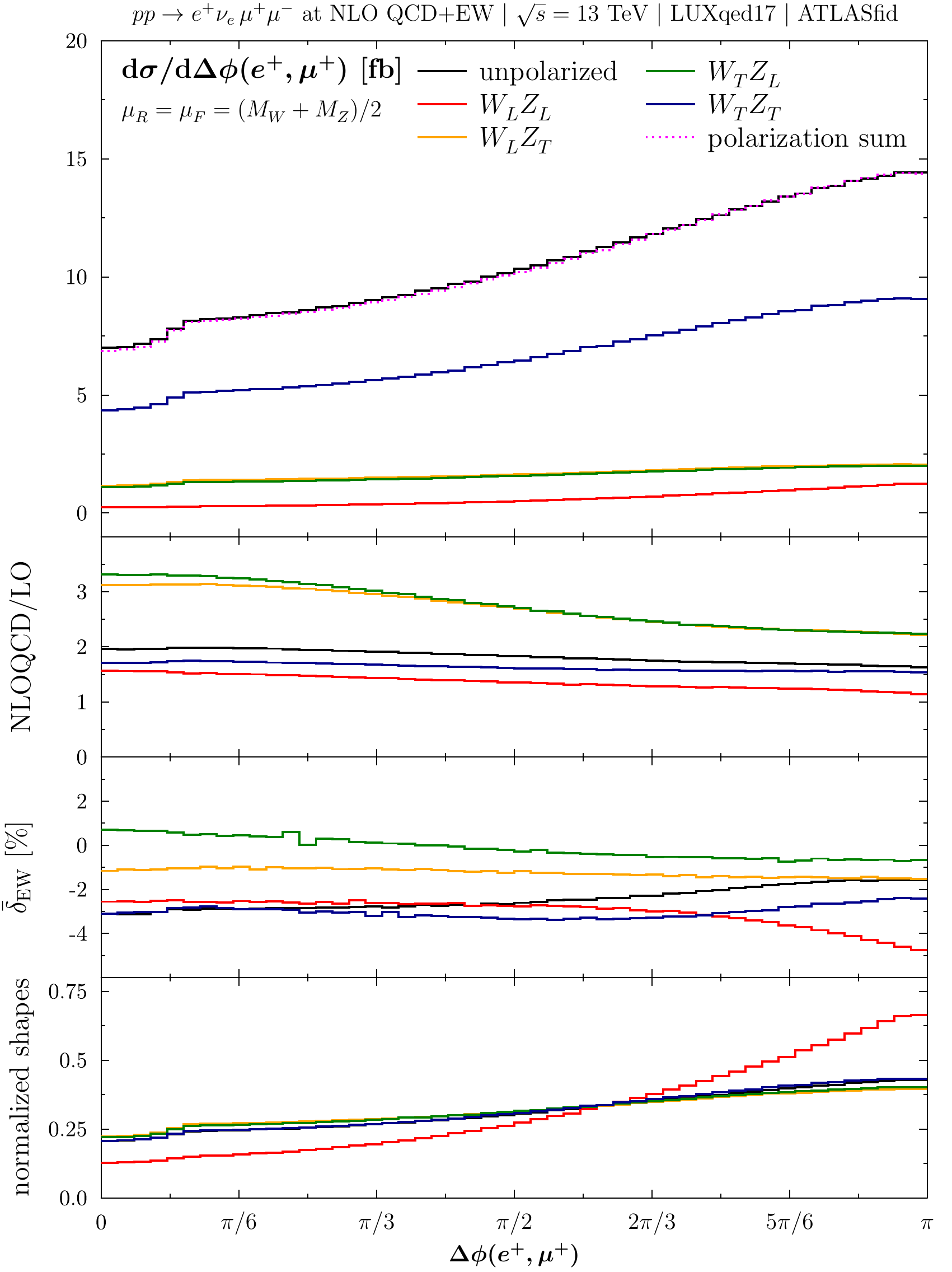}
  \end{tabular}
  \caption{Same as \fig{fig:dist_costheta} but for the azimuthal angles between the momenta of the 
electron and the muons, $\Delta\phi (e^+,\mu^-)$ (left) and $\Delta\phi (e^+,\mu^+)$ (right).}
  \label{fig:dist_phi}
\end{figure}

We next move to distributions in azimuthal angles, namely the angles between the momenta of the 
electron and the muons, $\Delta\phi (e^+,\mu^-)$ and $\Delta\phi (e^+,\mu^+)$. We obviously expect that 
different polarizations contribute differently to these observables. The results are provided in \fig{fig:dist_phi}, 
presented in the same format as the previous distributions. The first interesting thing to notice is that the $W_T^{}Z_L^{}$ and $W_L^{}Z_T^{}$ 
are the same in magnitude and in shape. They are only different in the EW corrections (middle-down panels), but these 
effects are too small to be visible in actual measurements. The shape of the $W_L^{}Z_L^{}$ is distinctly different from the other ones, 
hence this can be used as a discriminator to measure the $W_L^{}Z_L^{}$ component. The EW corrections are small, 
being from $-5$\% to $+1$\% for all polarizations and for both distributions.

\begin{figure}[t!]
  \centering
  \begin{tabular}{cc}
  \includegraphics[width=0.48\textwidth]{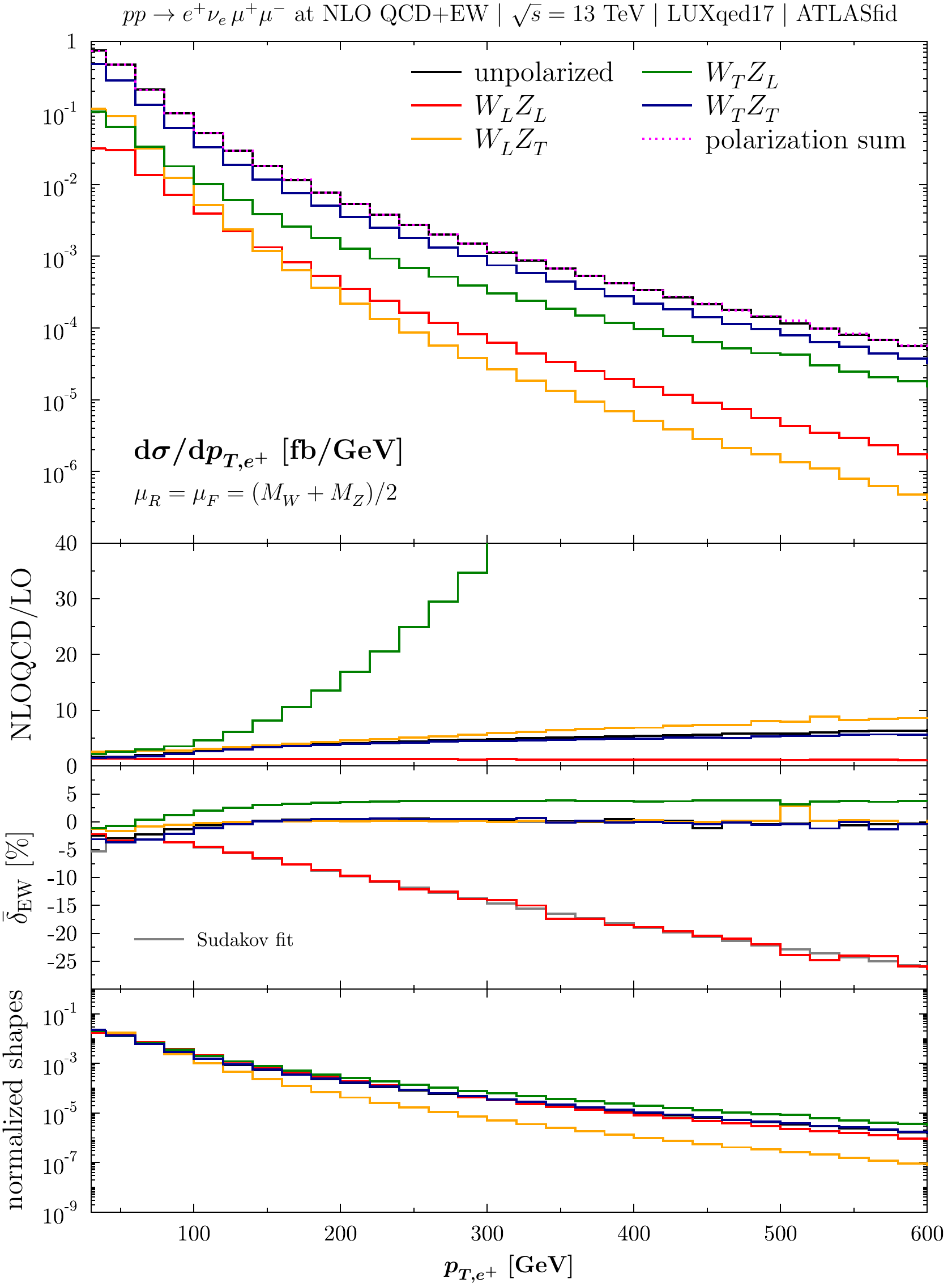}& 
  \includegraphics[width=0.48\textwidth]{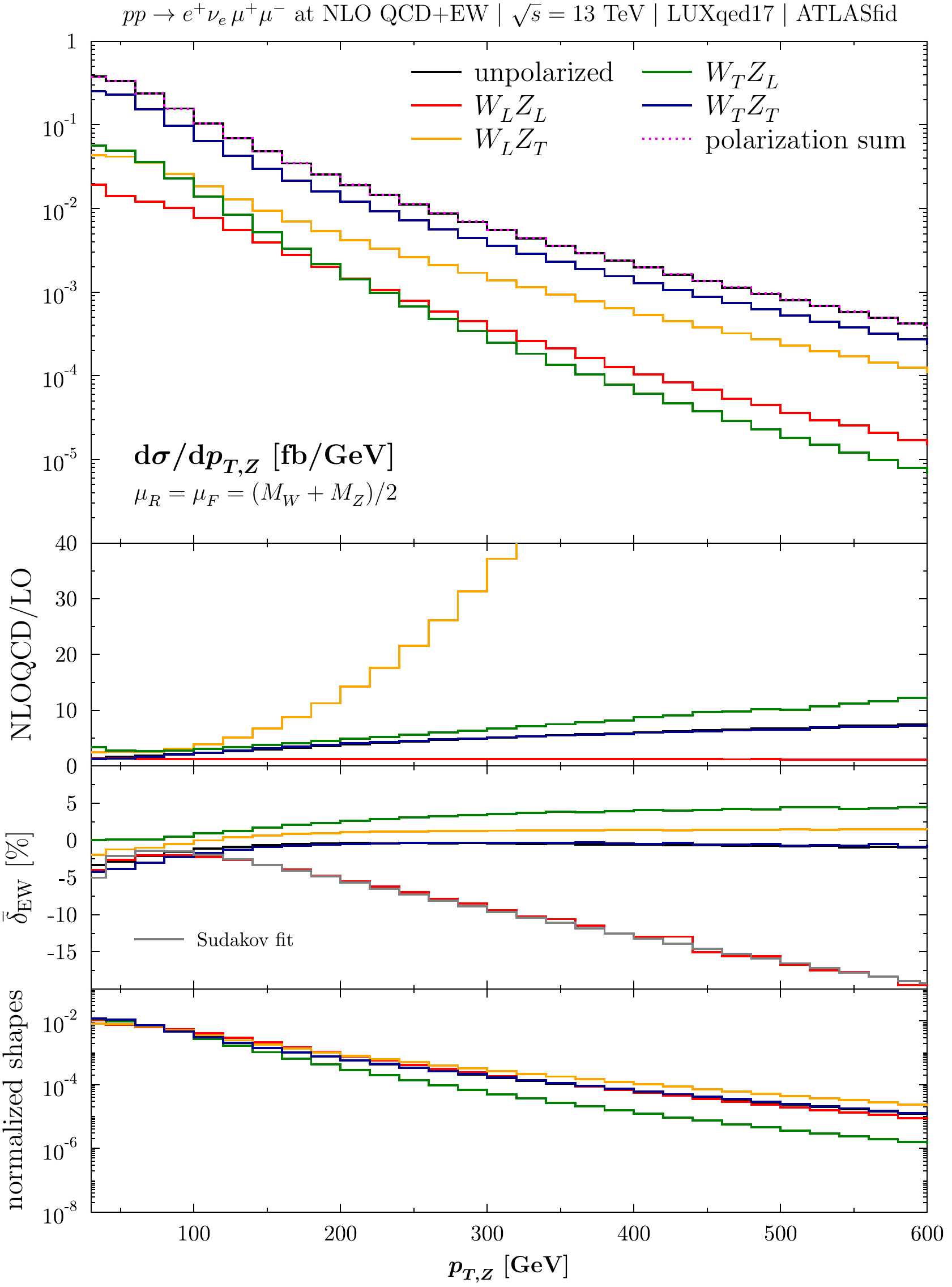}
  \end{tabular}
  \caption{Same as \fig{fig:dist_costheta} but for the transverse momentum of the 
electron (left) and the $Z$ boson (right). In the middle-down panel, the grey line is the Sudakov fit (see text) of the $W_L^{}Z_L^{}$ 
EW correction.}
  \label{fig:dist_pT}
\end{figure}
Finally, the transverse momentum distributions for the electron and the $Z$ boson are shown in \fig{fig:dist_pT}. 
Very unexpectedly, as opposed to the above angular distributions, 
the $W_L^{}Z_L^{}$ contributions are not smallest in both plots at large $p_T$. For the electron case, 
at large $p_{T,e}$, the $W_L^{}Z_L^{}$ and $W_L^{}Z_T^{}$ components fall fastest and become very small. They must vanish in the large $p_{T,e}$ limit, 
being equal to the Goldstone contribution, according to the equivalence theorem \cite{Willenbrock:1987xz}. At small $p_{T,e}$, 
the $W_L^{}Z_L^{}$ is smallest. With increasing $p_T$, the $W_L^{}Z_T^{}$ drops faster and becomes smallest at around $150$~GeV. 
Similar phenomenon happens for the $p_{T,Z}$ case: the $W_T^{}Z_L^{}$ becomes smaller than the $W_L^{}Z_L^{}$ at around $200$~GeV. 
Same kind of behavior was obtained in \cite{Denner:2021csi} (see Figs. 8 and 9 there) for the $ZZ$ process. 
To understand why the $W_L^{}Z_T^{}$ and $W_T^{}Z_L^{}$ contributions can be so small, it is interesting to look at the LO results 
(the reader can also see this from the big panels by removing the QCD corrections using the information in 
the NLO QCD/LO panels. The EW corrections are small and irrelevant here.). 
The picture at LO (not shown here) for the $p_{T,e}$ distribution reads: at small $p_{T,e}$ the $W_T^{}Z_L^{}$, $W_L^{}Z_T^{}$, $W_L^{}Z_L^{}$ 
are at the same order of magnitude; then with increasing momentum the $W_T^{}Z_L^{}$ and $W_L^{}Z_T^{}$ drop much faster and become 
much smaller than the $W_L^{}Z_L^{}$. We now take into account the QCD corrections. \fig{fig:dist_pT} (left) shows that 
the $W_T^{}Z_L^{}$ gets a huge correction, the $W_L^{}Z_T^{}$ a large correction, and the $W_L^{}Z_L^{}$ a small correction. This changes the hierarchy, 
making the $W_T^{}Z_L^{}$ largest and $W_L^{}Z_T^{}$ smallest at NLO QCD. Similar things happen in the $p_{T,Z}$ distribution with the $W_T^{}Z_L^{}$ and $W_L^{}Z_T^{}$ interchanged.  

The other important result is the magnitude of the EW corrections, which can be important for the interesting case of doubly longitudinal 
cross section. EW correction is about $-20$\% at $p_{T,e}\approx 450$~GeV, and is about $-10$\% at $p_{T,e}\approx 200$~GeV 
which is currently accessible at the LHC (see \refs{ATLAS:2019bsc,CMS:2021icx}). For the $p_{T,Z}$ distribution, the 
corrections are significantly smaller. This correction originates from the negative Sudakov corrections in the virtual contribution. 
It can therefore be fitted using the single and double Sudakov logarithms. 
Our fit yields
\begin{align}
\bar{\delta}^\text{fit,e}_{\text{EW}}&=-0.034 \left[1 + 0.7 \log\left(\fr{p_{T,e}}{M_W}\right) + 1.3 \log^2\left(\fr{p_{T,e}}{M_W}\right) \right],\\
\bar{\delta}^\text{fit,Z}_{\text{EW}}&=-0.015 \left[1 + \log\left(\fr{p_{T,Z}}{M_Z}\right) + 2.8 \log^2\left(\fr{p_{T,Z}}{M_Z}\right) \right],
\end{align}
where a constant term has been added in the fit to account for the low energy regime. These fits are shown in the plots (grey line), showing excellent agreement with the exact values. For the other polarizations, EW corrections are smaller than $5$\%, hence can be neglected.   
\section{Conclusions}
\label{sect:conclusion}
We have presented, for the first time, the NLO EW corrections to the doubly-polarized cross sections of the process 
$pp \to W^+ Z \to e^+ \nu_e \mu^+ \mu^- + X$ at the LHC. The results are of direct consequences to the measurements of 
double-polarization signals in the $WZ$ production channel at the LHC. To be as close as possible to the current experimental 
setup, the ATLAS fiducial cuts have been used and the polarization signals are defined in the $WZ$ c.m.s as implemented in the 
latest polarization measurement by ATLAS \cite{ATLAS:2019bsc}. 

For completeness and putting everyone in the same footing, we have re-calculated the known NLO QCD corrections and obtained good agreement with the results of Denner and Pelliccioli \cite{Denner:2020eck}. The QCD corrections are then combined with the EW ones to obtain the full NLO QCD+EW results. 

We found that the impact of EW corrections on the integrated polarized cross sections is negligible, 
being smaller than $3$\% (relative to the NLO QCD results) for all polarizations. 
For angular distributions ($\cos\theta^{WZ}_{e^+}$, $\cos\theta^{WZ}_{\mu^-}$, 
$\Delta\phi (e^+,\mu^-)$, and $\Delta\phi (e^+,\mu^+)$), the EW corrections are also very small, 
being smaller than $5$\% across the full ranges. For transverse momentum distributions ($p_{T,e}$ and $p_{T,Z}$), the EW corrections are found to be sizable only in the doubly longitudinal cross section. At 
the current LHC accessible range of $p_{T,e}\approx 200$~GeV, the correction is about $-10$\%. 
The magnitude of the correction increases rapidly with $p_{T}$. The shape of this correction 
can be excellently fitted using the single and double Sudakov logarithms.  

\acknowledgments

This research is funded by the Vietnam National Foundation for Science and
Technology Development (NAFOSTED) under grant number
103.01-2020.17.



\providecommand{\href}[2]{#2}\begingroup\raggedright\endgroup

\end{document}